\documentclass[%
 reprint,
 amsmath,amssymb,
 aps,
]{revtex4-2}
\usepackage{amsmath}
\usepackage{amssymb}
\usepackage{mathtools}
\usepackage{float}
\usepackage{graphicx,graphics,color}
\usepackage{stix}
\usepackage{graphicx}
\usepackage{dcolumn}
\usepackage{bm}
\let\vec\mathbf

\begin{document}

 \author{Chen-How Huang}
\affiliation{Donostia International Physics Center (DIPC), 20018 Donostia-San Sebastian, Spain}
\title{ A Microscopic Description for Two Body Loss in Cold Atoms Near Feshbach Resonances with Strong Spontaneous Emission}
 \begin{abstract}
We study the two body loss dynamics of fermionic cold atoms near $s$- and $p$-wave Feshbach resonances with a microscopic Keldysh path integral formalism and compare the result to the macroscopic phenomenological loss rate equation. The microscopic loss rate equation is an integral-differential equation of the momentum distribution that depends on the functional form of the loss rate coefficient. For $s$-wave resonance, the microscopic theory yields the same result as the phenomenological equation. However, the calculation of $p$-wave resonance shows a discrepancy between the two descriptions for an quantum-degenerate gas. This discrepancy originates from the functional form of the loss coefficient which is associated with the microscopic loss mechanism of the two body loss and is neglected in the phenomenological equations where the coefficient is typically a constant. We find the discrepancy between the microscopic theory and the phenomenological description is smeared by thermal average at high temperature, $T\gtrsim T_F$ where $T_F$ is the Fermi temperature.

\end{abstract}
\maketitle

\section{Introduction}

\textcolor{black}{
  Dilute cold gas is an important experimental platform due to its flexibility of controlling the system parameters  for quantum simulations of many-body systems. The interaction of cold atoms can be controlled by Feshbach resonances\cite{
RevModPhys.82.1225_CC,DUINE2004115_review,
TIMMERMANS1999199_review,RevModPhys.78.483_review,RevModPhys.78.1311_review} using either magnetic or optical tuning methods where the scattering in the ground state is modulated by coupling the scattering channels in the ground state to excited molecular states.
 }
Near Feshbach resonance, two body loss is an inevitable phenomenon since the decaying of molecular states to two particles in the ground state causes a large gain in the kinetic energies that allows the particles to escape from the trap. The two body loss dynamics is typically described by a phenomenological equation of the total density\cite{ Bohn_PhysRevA.56.1486, PhysRevA.95.012708_Braaten,PhysRevLett.105.050405_OFR0,Blatt_PhysRevLett_2011,Ciurylo_PhysRevA_2004,enomoto_PhysRevLett.101.203201,TheisDenschlag2004,PhysRevA.82.062704_pwave,Yamazaki_PhysRevA.87.010704},
\begin{align}\label{eq:phen}
\frac{dn_c}{dt}=-\gamma  n_c^2, 
\end{align}
where the loss dynamics is described by the evolution of the macroscopic total particle density, $n_c$, and a single parameter $\gamma$. The coefficient is typically assumed to be a constant, which may neglect the microscopic process of the two body loss  mechanism.
For instance, recent studies\cite{PhysRevResearch.5.043192_CH,Rossini_PhysRevA.103.L060201} using different theoretical approaches on hard-core Bosons in one dimensional optical lattices with photo-associated two body loss\cite{Tomita_PhysRevA_2019,Syassen_science_2008} shows that the two body loss equation is a differential-integral equation of the instantaneous momentum distribution, and the loss coefficient depends on the geometry of the optical lattice and the microscopic mechanism of two body loss in the system. The phenomenological description is shown to be a high-temperature limit of the microscopic loss rate equation\cite{PhysRevResearch.5.043192_CH}.
Therefore, it is essential to check if there are discrepancy between the macroscopic phenomenological equation and the microscopic loss rate equation for cold atoms near Feshbach resonances.

In this article, we study the two body loss dynamics for cold atoms near Feshbach resonance. We describe the resonance scheme using a two channel Hamiltonian and derived the microscopic loss rate equation using Keldysh path integral\cite{kamenev_2011,Sieberer_IOP_2016} and second-order cumulant expansion\cite{Kubo1980}. The application of these theoretical techniques to describe particle losses is explained in \cite{PhysRevResearch.5.043192_CH}. We find the phenomenological equation is a good description of the loss dynamics for $s$-wave resonance or a thermal gas. However, for $p$-wave resonances, it is more accurate to describe the dynamics using the microscopic  loss rate equation of the momentum distribution and the phenomenological equation is a high temperature limit of the microscopic description.

This article is organized as the following. In section~\ref{sec:model} we introduce the model Hamiltonian and derive the microscopic loss rate equation using Keldysh path integral formalism. In section~\ref{sec:2bdloss} we discuss and compare the loss dynamics for $s$-wave resonance obtained from our microscopic theory and the phenomenological equation. In section~\ref{sec:$p$-wave}, we calculate the dynamics for $p$-wave resonance, and discuss the difference between the microscopic and phenomenological equation. Section~\ref{sec:fin} is the conclusion. Throughout this paper, we use $\hbar=1$.
\begin{figure}[h]
\includegraphics[width=0.95\columnwidth]{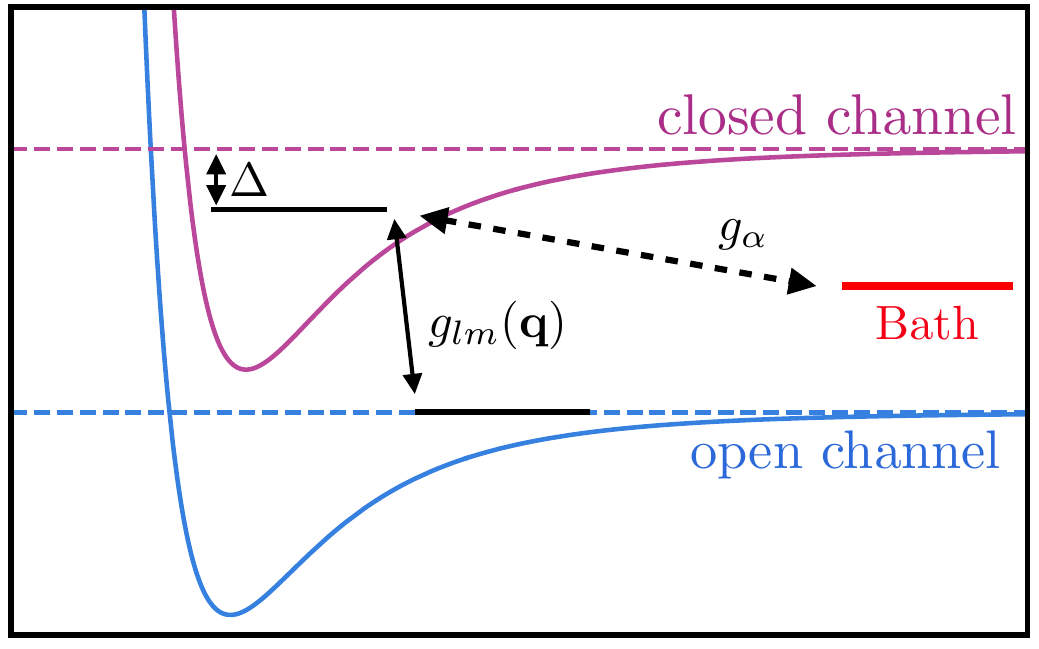}
\center
\caption{Scheme of a Feshbach resonance: The system  consists  of cold atoms in the ground state open channel interacting via an $s$-wave potential.  The colliding pairs of atoms in the open channel are coupled to the molecular state in the closed channel. This state has a one-body coupling parametrized by $g_{\alpha}$ to a bath to which it can be lost.}
\end{figure}

\section{two channel model for dilute fermionic cold atoms near Feshbach resonance}\label{sec:model}
 In this section we describe a fermionic cold atom system near a Feshbach resonance associated with the $l$-wave  scattering channel by means of the following model.  Here $l=0,1$ for $s,p$-wave. To illustrate our idea, we use a two-component system as the minimum model.
\begin{align}
H_{c}&= \sum_{\vec{k}\sigma} \epsilon_{\vec{k}} c^\dag_{\vec{k}\sigma} c_{\vec{k}\sigma} + H_{\mathrm{int}},\\
H_{\mathrm{int}} &= \frac{U}{\Omega}\sum_{\vec{pkq},\sigma\sigma'} Q_{\sigma\sigma'}c^{\dag}_{\vec{p}\sigma}
c^{\dag}_{\vec{k}\sigma^{\prime}} c_{\vec{k-q}\sigma^{\prime}}c_{\vec{p+q}\sigma},\\
H^l_{ca} &= \frac{S(t)}{\sqrt{2\Omega}}\sum_{\vec{pq}\sigma\sigma'} \sum_{m=-l}^l    Q_{\sigma\sigma'} g_{lm}(\vec{q})  a^{\dag}_{m\vec{p}  } c_{\vec{p}-\vec{q},\sigma} c_{\vec{q},\sigma^{\prime}}+h.c.,\\
H^l_a&= \sum_{\vec{q} }\sum_{m=-l}^l  (\epsilon^a_{\vec{q}} + S'(t)\Delta) a^\dag_{m\vec{q}  }a_{m\vec{q}  },\\
H^l_{aB} &=    \sum_{\vec{q},\alpha} \sum_{m=-l}^l    \left[ g_{\alpha} b^{\dag}_{\vec{q}\alpha}a_{m\vec{q}  } +  g^*_{\alpha} a^\dag_{m\vec{q} }b_{\vec{q},\alpha} \right],\\
H_\text{B}&=  \sum_{\vec{q},\alpha} \omega_{\vec{q}\alpha}\: b^\dag_{\vec{q}\alpha}b_{\vec{q}\alpha},
\end{align}
Here $c_{\vec{k}\sigma}$ and $c^\dag_{\vec{k}\sigma}$ are the fermionic operators of the ground state atoms with spin $\sigma=\uparrow,\downarrow$ and momentum $\vec{k}$ in the open channel satisfying the anticommutator relation $[c_{\vec{k}\sigma},c_{\vec{k'},\sigma'}]_+=\delta(\vec{k}-\vec{k'})\delta_{\sigma,\sigma'}$. They  interacts via $s$-wave background interaction $H_{\mathrm{int}}$. $a_{m\vec{k}}$ and $a^\dag_{m\vec{k}}$ satisfing $[a_{m\vec{k}},a^\dag_{m'\vec{k'}}]_-=\delta(\vec{k}-\vec{k'})\delta_{m,m'}$ are the bosonic operators of the molecular state in the closed channel with the projection of angular momentum in the $l$-wave scattering channel, $m=-l,\cdots, l$, momentum $\vec{k}$. The molecules are lost by coupling to the bosonic field of the bath particles with eigenmodes $\alpha$ and momentum $\vec{k}$ denoted as $b_{\vec{k}\alpha},b^\dag_{\vec{k}\alpha}$. $S(t)$ and $S'(t)$ describe the time dependence of the laser coupling and detuning. Here, we will consider $S(t)=\Theta(t)$ and $S'(t)=1$ which describes the dynamics of turning on the laser suddenly.

The $l$-wave resonance is applied by a laser coupling between the bosonic molecular level in the closed channel and a pair of ground state fermions with different spins. Here $Q_{\sigma\sigma'}=
1-\delta_{\sigma,\sigma'}$, and the coupling $g_{lm}(\vec{q})=gq^lY_{lm}(\vec{\hat{q}})$ where $Y_{lm}(\vec{\hat{q}})$ is the spherical harmonic functions.
 This expression arises from $l$-wave symmetry of the two-body scattering wave functions  which comes from the matrix element, $_{00}\langle e|V_{las}| g\rangle_{lm}$, of the laser-induced transition between two-particles in the $l$-wave collision channel of ground state and the $s$-wave channel of the molecular excited states associated with $l$-wave resonance.
 We note, the momentum dependence can be more complicate and depend on the detail of the scattering potential in the molecular state. However, since our goal is not to address a specific experimental situation but provide a  general discussion of the two body loss dynamics in Feshbach resonances. We pick the $q^l$ dependence from the scattering wave function of free particles.
\begin{align}
&_{00}\langle e |V_{las}| g \rangle_{lm}\notag\\
&\simeq\langle\gamma_e|V_{las}|\gamma_g\rangle\int_{0}^{\min[1/k_F,\Lambda]} j_0(qr) Y^*_{00}(\vec{\hat{q}}) j_l(qr) Y_{lm}(\vec{\hat{q}})  dr,\\
&=   gq^lY_{lm}(\vec{\hat{q}}),
 \end{align}
 where $\gamma_g$ and $\gamma_e$ labels the quantum numbers of the non-spatial part of the scattering wave functions in the ground and excited states. The $q^l$ dependence arises from the spherical Bessel function $j_l(qr)\propto (qr)^l $ at $qr \ll 1$. 
This expression is valid if the range of the interaction, $R$, satisfies $R/\Lambda \ll 1$ and $n R^3\ll 1$ where $\Lambda$ is the thermal de Broglie length and $n$ is the density of the system. We also assume the scattering length of the background s-wave interaction, $H_{\mathrm{int}}$, satisfies $k_Fa_{\text{bg}}\ll 1 $ such that we can neglect the vertex corrections to the coupling\cite{PhysRevLett.115.135303_pwave_2,PhysRevLett.99.190406_pwave_1} and their contribution to the loss dynamics is negligible in our leading order perturbation calculation.

\textcolor{black}{
This model amounts for a two channel description of Feshbach resonance of a weakly interacting dilute gas. The two body loss dynamics for $s$-wave resonance has been studied using a single channel model\cite{PhysRevA.95.012708_Braaten}, and the same loss rate equation can be derived with our two-channel model(e.g. section~\ref{sec:2bdloss}). The momentum dependence in $g_{lm}(\vec{q})$ emerges for $l>0$ and leads to a distinct behavior from the phenomenological equation(e.g. section~\ref{sec:$p$-wave}  for detail discussion).}
The two body loss of the system particles is described as the one-body loss of the intermediate molecular state, which is coupled to the Markovian bath via $H^l_\text{aB}$ whose eigenmodes are the bosonic operators $b^{\dag}_{\vec{q}\alpha}, b_{\vec{q}\alpha}$.
To proceed, we write down the Keldysh generating funtional and integrate out the bath field using second order cumulant expansion following the treatment for a Markovian bath in \cite{PhysRevResearch.5.043192_CH},
\begin{align}
Z &= \int D[\bar{b}b,\bar{a}a,\bar{c}c] \, e^{i S^l},\\
&=Z_b\int D[\bar{a}a,\bar{c}c] \, e^{i S^l_{\text{eff}}},
\end{align}
where $S^l$ is the Keldysh action of the full Hamiltonian, and $S^l_{\text{eff}}$ is the following Keldysh action with one-body loss in the molecular field and is derived by integrating out the bath field,
\begin{align}
&S^l_{\text{eff}}=S^l_c + S^l_{a,\mathrm{eff}},\\
&S^l_c= \int dt\, \sigma^3_{rs}  \biggl\{ \left[\sum_{\vec{k}} \bar{c}_{\vec{k}\sigma r}\left( i\partial_{t}-\epsilon_{\vec{k}}  \right) c_{\vec{k}\sigma s}  - H_{\mathrm{int}}\right]\notag\\
       &-  \frac{\Theta(t)}{\sqrt{
    \Omega}}\sum_{\vec{p}\vec{q},m,\sigma \sigma^{\prime}}    Q_{\sigma\sigma'}\biggl[ g_{lm}(\vec{q}) \bar{a}_{m\vec{p}  r} c_{\vec{p}-\vec{q}\sigma s}c_{\vec{q}\sigma^{\prime}s} \notag\\ & \qquad\qquad+ g^*_{lm}(\vec{q})a_{m\vec{p} r} \bar{c}_{\vec{p}-\vec{q}\sigma s} \bar{c}_{\vec{q}\sigma^{\prime} s} 
    \biggr] \biggr\},\notag \\
& S^l_{a,\mathrm{eff}} =    \sum_{\vec{q}} \sum_m \int dt  dt^{\prime}\, 
\bar{a}_{m\vec{q} r} G^{-1}_{\vec{q},rs}(t,t^{\prime}) a_{m\vec{q} s}\,,
\label{eq:Sm'}
\end{align} 
where
\begin{multline} 
G^{-1}_{\vec{q},rs}(t,t^{\prime}) = \delta(t-t^{\prime}) \left[ 
\sigma^{3}_{rs}  \left(i\partial_{t^{\prime}} - \epsilon^a_{\vec{q}} - \Delta \right) \right. \\
\left. + i \sigma^0_{rs}\gamma /2 - i\sigma^{-}_{rs}\gamma  \right]. \label{eq:ginv}
\end{multline}
 The label $r,s = \pm$ represent the time ordered and the anti-time ordered segment of the Keldysh contour. $\sigma^3_{rs}=r\delta_{r,s}$ and $\sigma^0_{rs}=\delta_{r,s}$ are the Pauli matrix and identity matrix. In Eq.~\eqref{eq:Sm'}, in order to lighten the equations, we have adopted the convention that summation over
repeated indices $r,s =\pm$ is implied. 
In the calculations below, we assume that  $ g_{lm}\ll \min\{\Delta,\gamma\}$, where  $\Delta$ is the detuning from the excited molecular state and $\gamma$ is the one-body loss rate of the excited molecular state which relates to the spontaneous emission rate of the molecular states by $\gamma=2\pi \Gamma_{spon}$. In this limit, the loss of the molecular state is due to spontaneous emission. The contribution of stimulated emissions is relatively small. 

Next, we integrate out the molecular field, $\bar{a}_{\vec{q} }, a_{\vec{q}}$ using the second order cumulant expansion. The Green's function for the molecular fields is derived by inversion of the matrix $G^{-1}_{\vec{q},rs}(t,t^{\prime})$. Within the limit, $g_{lm}\ll \min[\Delta,\gamma]$,  we treat the molecular field as a Markovian bath which amounts to
neglecting the term involving the time derivative, $i\partial_t$. A detailed reasoning for this approximation and derivation of the Keldysh action was explained in \cite{PhysRevResearch.5.043192_CH}. We get the Markovian correlations for the molecular field of the form:
\begin{multline}
G_{\vec{q},rs}(t,t^{\prime}) \simeq  -\frac{\delta(t-t^{\prime})}{(\epsilon_{\vec{q}}+\Delta)^2+ \gamma^2 /4}\\
\times \left[ \sigma^{3}_{rs}  \left(\epsilon^a_{\vec{q}} + \Delta \right)  + \tfrac{i}{2}\left( \sigma^0_{rs} + \sigma^{-}_{rs} \right) \gamma  \right].
\end{multline}
 Using the above expression, we obtain the  following effective Keldysh action:
\begin{align}\label{eq:fesh2}
 S^l_{\text{eff}}&=S_0+\mathcal{L},\\
 S_0&=\sum_{\vec{k},\sigma} \int dt  \sigma^3_{rs} \bar{c}_{\vec{k}\sigma r}\left( i\partial_{t}  - \epsilon_{\vec{k}} \right) c_{\vec{k}\sigma s}-\sigma^3_{rs}H_{int},\\
\mathcal{L}&= \int dt\biggl\{
- \frac{\Theta(t) }{2\Omega}\sum_{\vec{p}\vec{k}\vec{q}} \sum_{\sigma \sigma^{\prime} }Q_{\sigma\sigma'}
\biggl[ \sigma^3_{rs}   \delta U_l (\vec{p},\vec{k},\vec{q}) \notag\\
 &\qquad\qquad\quad - i\sigma^0_{rs} \gamma_l^{\prime}(\vec{p},\vec{k},\vec{q}) \biggr]  \bar{c}_{\vec{p}\sigma r}\bar{c}_{\vec{k}\sigma^{\prime} r}c_{\vec{k+q} \sigma^{\prime}  s}
c_{\vec{p-q},\sigma  ,s} \biggr\}\notag \\
-i &   \int dt  \sum_{\vec{p}\vec{k}\vec{q}} \sum_{\sigma \sigma^{\prime} }  
\frac{\Theta(t)\gamma_l^{\prime}(\vec{p},\vec{k},\vec{q})Q_{\sigma\sigma'}}{\Omega}  \bar{c}_{\vec{p}\sigma -}\bar{c}_{\vec{k}\sigma^{\prime} -}
c_{\vec{k+q}\sigma^{\prime}+} c_{\vec{p-q}\sigma+},
\end{align}
where $\delta U_l(\vec{p},\vec{k},\vec{q})$ is the correction to the interaction: 
\begin{align}
       \delta U_l(\vec{p},\vec{k},\vec{q}) =-2\sum_m  
    g^*_{lm}(\vec{p}-\vec{k}) g_{lm}(\vec{p}-\vec{k}-2\vec{q})   
    \frac{\Delta_{\vec{p}+\vec{k}}  }{\Delta^2_{\vec{p}+\vec{k}}+ \tfrac{\gamma^2}{4}},
\end{align}
and  
\begin{align}
\gamma_l^{\prime}(\vec{p},\vec{k},\vec{q})=\sum_mg^*_{lm}(\vec{p}-\vec{k})g_{lm}(\vec{p}-\vec{k}-2\vec{q})
  \frac{\gamma   }{\Delta^2_{\vec{p}+\vec{k}}+\frac{\gamma^2 }{4}},  
\end{align}
is the effective two body loss rate in the limit of strong spontaneous loss on the intermediate molecular states~\cite{Thies_PhysRevLett.93.123001,Ciurylo_PhysRevA_2004,enomoto_PhysRevLett.101.203201,Napolitano_PhysRevLett.73.1352,Yamazaki_PhysRevA.87.010704,Kim_PhysRevA.94.042703,Bohn_PhysRevA.56.1486}. In the above expressions $\Delta_{\vec{p}}= \epsilon^a_{\vec{p}}+\Delta $ is the energy of the excited molecular state with total momentum $\vec{p}$. In the following calculations, we will assume the kinetic energy of the molecule is much smaller than the detuning, $ \epsilon^a_{\vec{p}}\ll\Delta$, such that $\Delta_{\vec{p}}\simeq\Delta$. Notice that both the corrections to the interaction and the effective loss rate are perturbatively small in the limit, $ g_{lm} \ll \min\{\Delta, \gamma\}$, of interest here. Hence, perturbation theory to leading order in $\gamma_l^{\prime}$ yields the following rate equation for an quantum degenerate gas. The detailed derivation for the perturbation calculation is explained in appendix~\ref{app:A}.
\begin{align}\label{eq:micro}
\frac{dn_\sigma(\vec{p},t)}{dt}&= -\frac{2}{\Omega^2}\sum_{\vec{k}}\sum_{  \sigma',\sigma\neq\sigma'} 
  \gamma'_{l}(\vec{p},\vec{k},\vec{0}) n_\sigma(\vec{p},t)n_{\sigma'}(\vec{k},t),
\end{align}
where $n_{\vec{p},\sigma}(t)$ is the instantaneous momentum distribution for spin $\sigma$ particles with momentum $\vec{p}$, and $\gamma_l'(\vec{p},\vec{k},\vec{0})$ is,
\begin{align}
\gamma_l'(\vec{p},\vec{k},\vec{0})&=\sum_m |g_{lm}(\vec{p-k})|^2\frac{\gamma }{\Delta_{\vec{p+k}}^2+ \gamma^2/4},\\
&=\frac{1}{|\vec{p-k}|  }\frac{\Gamma_{spon}\Gamma_{sti;l}(\vec{p-k})}{(2\pi\Delta)^2+ \Gamma^2_{spon}/4},
\end{align}
where, in the last line, we have related the coupling to the stimulated width, $\Gamma_{sti;l}(\vec{q})/2\pi q=g^2 q^{2l} \sum_m |Y_{lm}(\vec{\hat{q}})|^2$ by Wigner threshold law. The $q$ dependence in the denominate of $\Gamma_{sti;l}(\vec{q})/q$ arises from the density of state in 3D $\propto \sqrt{E}\sim q$. 
 Eq.~\eqref{eq:micro} infers that the dynamics is determined by the functional form of the loss coefficient and the initial momentum distribution. Since our goal is to benchmark the validity of the phenomenological equation and our microscopic theory, we will consider an initial state of Fermi-Dirac distributions.

The phenomenological loss coefficient for a thermal gas can be obtained by replacing the loss coefficient with its thermal average~\cite{RevModPhys.82.1225, Ciurylo_PhysRevA_2004,Blatt_PhysRevLett_2011},
\begin{align}
\gamma_{T}&=\langle \gamma_{l}^{\prime}(\vec{p},\vec{k},\vec{0}) \rangle_T,\\
&=\int \gamma_{l}^{\prime}(\vec{p},\vec{k},\vec{0}) f_M(\vec{p},T) f_M(\vec{k},T) d^3\vec{p} d^3\vec{k}.
\end{align}
Here $f_M(\vec{p},T)=(2\pi m k_B T )^{-3/2}\exp(-p^2/2m k_B T)$ denotes the  normalized Maxwell distribution at temperature $T$ for particles with mass $m$. Using the average coefficient, the rate equation can be approximated to,
\begin{align}\label{eq:ph}
\frac{d n_\sigma(\vec{p},t)}{dt}&= -  \frac{2\gamma_{T}}{\Omega^2}\sum_{\vec{k}}\sum_{\sigma',\sigma\neq\sigma' } n_\sigma(\vec{p},t)n_{\sigma'}(\vec{k},t),
\end{align}
which is the phenomenological two body loss rate equation for a thermal gas~\cite{RevModPhys.82.1225,Ciurylo_PhysRevA_2004,enomoto_PhysRevLett.101.203201,Blatt_PhysRevLett_2011,Yamazaki_PhysRevA.87.010704}. For an initially equally populated state of $2$-components, the phenomenological equation reduced to a differential equation of the total density, $n_c(t)=\sum_{\sigma,\vec{p}}n_{\sigma}(\vec{p},t)$, expected in eq~\eqref{eq:phen}, 
\begin{align}\label{eq:ph}
\frac{d n_c(t)}{dt}&= -\gamma_T \: n_c^2(t),
\end{align}
\section{ s-wave two body loss dynamics for two component fermions}\label{sec:2bdloss}
\begin{figure}[t]
\includegraphics[width=\columnwidth]{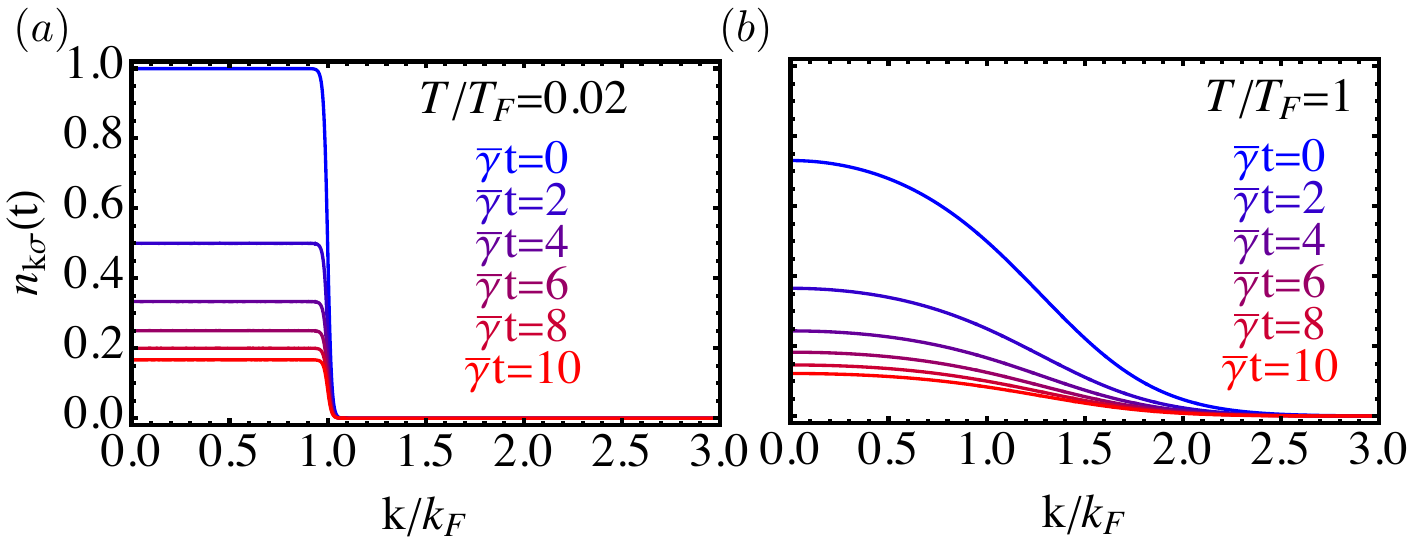}
\center
\caption{    Evolution of the momentum distribution at initial temperatures (a) $T=0.02T_F$ and (b) $T=T_F$ for an equally populated two component system near $s$-wave resonance. $k_F$  is determined by the Fermi momentum at zero temperature limit. Here $\bar{\gamma} = 2\gamma_0'$ in eq.\eqref{eq:$s$-wave}.}
\label{fig:2bdloss}
\end{figure}
For $s$-wave resonance, $l=0$, the loss rate coefficient $\gamma'_0$ does not depend on momentum. The loss rate equation reads,
\begin{align}\label{eq:$s$-wave}
\frac{d n_\sigma(\vec{p},t)}{dt}&= - \frac{2\gamma'_0}{\Omega^2}\sum_{\sigma',\sigma\neq\sigma'}n_\sigma(\vec{p},t)\left[\sum_\vec{k}n_{\sigma'}(\vec{k},t)\right].
\end{align}
and is the same as the phenomenological equation by replacing $\gamma'_0=\gamma_T$. We note the decay, $dn_{\sigma}(p,t)/dt\propto n_{\sigma}(p,t)$, this means,  for a thermal gas, the low momentum particles experiences a stronger depletion since their higher occupation. Indeed, figure~\ref{fig:2bdloss}(b) shows the evolution of momentum distribution at high temperatures. The decrease in low momentum parts in the momentum distribution is much intenser than the high energy tails. On the contrary, panel (a)  shows nearly the same decay rate for the low momentum part since the initial occupation at low momentum is nearly the same below the Fermi level at low temperatures.  

The evolution of a macroscopic observable of the following form, $O_\sigma(t;f) = \sum_\vec{p} f(\vec{p}) n_\sigma(\vec{p},t)$, can be derived,
\begin{align}\label{eq:$s$-wave-special}
\frac{d O_\sigma(t;f)}{dt}&= - \frac{2\gamma'_0}{\Omega^2}\sum_{\sigma',\sigma\neq\sigma'}O_\sigma(t;f)\left[\sum_\vec{k}n_{\sigma'}(\vec{k},t)\right].
\end{align}
Therefore, the dynamics of $O_\sigma(t;f)/O_\sigma(t=0;f)$ is universal for any functional form of $f(\vec{p})$ and only depends on the total particle density and the loss rate, $\gamma'_0$. From the equation, we expect no difference between the phenomenological description and the microscopic theory since $\gamma'_0$ is a constant. 
\textcolor{black}{ These resemblances originate from our calculation where the interactions $H_{\mathrm{int}}$ and $\delta U$ are not included in the leading order perturbative expansion. Including these interaction leads to a vertex correction to the coupling between the molecules and atoms\cite{PhysRevA.71.063614_stoof}. The momentum dependence enters in the loss coefficients and may result in the difference between the microscopic and phenomenological theory. However, in the parameter region of our discussion, this effect is weak.} 

\section{ p-wave two body loss dynamics }\label{sec:$p$-wave}
\begin{figure}[t]
\includegraphics[width=0.99\columnwidth]{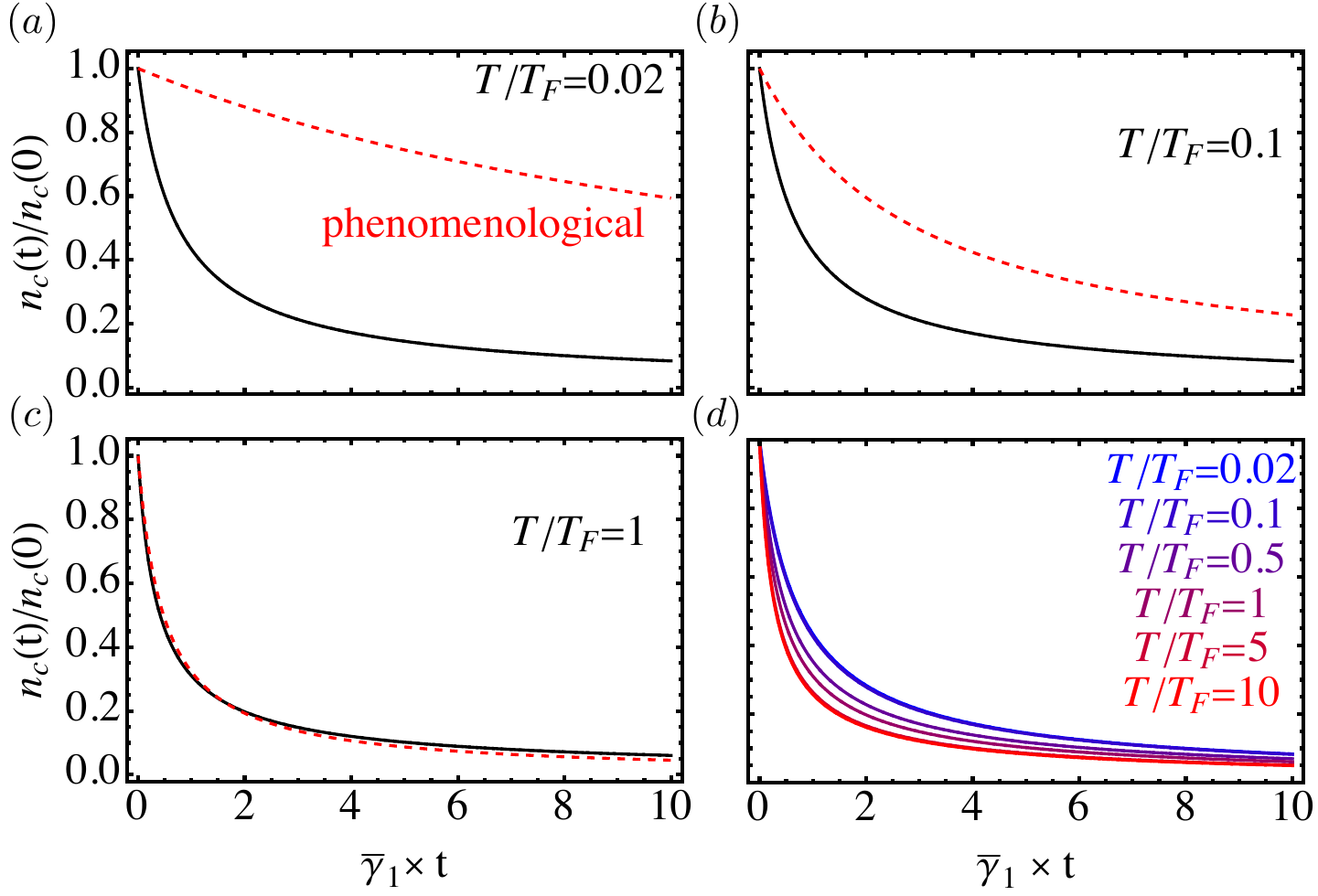}
\center
\caption{ (a), (b) and (c) are the Evolution of the total particle density near $p$-wave resonance at different initial temperatures for an equally populated two component system. A huge disagreement to the phenomenological result(red dashed lines) is observed at low temperatures. The difference decreases as we increase the temperature to $T\simeq T_F$. (d) Shows the evolution of the microscopic theory with different initial temperatures. The curves from above to below correspond to low($T/T_F=0.02$) to high($T/T_F=10$) initial temperatures. The decay of density is stronger for systems with higher initial temperatures.  }\label{fig:2bdloss_p}
\end{figure}
\begin{figure}[t]
\includegraphics[width=0.99\columnwidth]{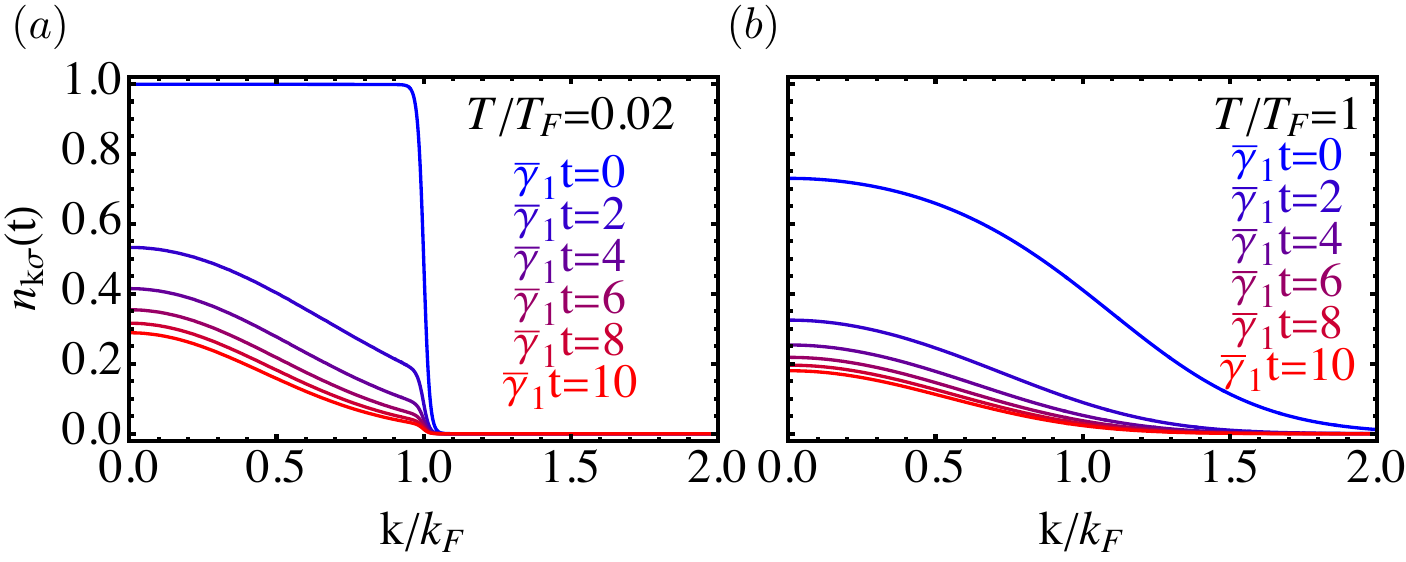}
\center
\caption{ Evolution of the momentum distributions near $p$-wave resonance of equally populated systems for (a) $T=0.02T_F$ and (b) $T=T_F$. $k_F$  is determined by the Fermi momentum at zero temperature limit. }\label{fig:2bdloss_p_m}
\end{figure}

For $l=1$, the momentum dependence of $p$-wave coupling causes discrepancies between the microscopic equation and the phenomenological result. Indeed, figure~\ref{fig:2bdloss_p}(a) ,(b) and (c) are the comparison to the phenomenological equation for the evolution of total particle density at different initial temperatures with $p$-wave resonance of an equally populated two component system. A huge difference is observed between the microscopic theory and the phenomenological result for temperatures $T\ll T_F$. This discrepancy becomes smaller as we reach higher temperatures, $T\gtrsim T_F$, in figure~\ref{fig:2bdloss_p}(c).
Panel (d) shows the evolution of density with different initial temperatures. Starting from the initial states with the same particle densities, the system with a higher temperature shows a stronger decay in the particle density. To understand this, we consider the evolution of the momentum distribution,
\begin{align}\label{eq:micro_eq}
&\frac{dn_{\sigma}(\vec{k},t)}{dt}= -\frac{2}{\Omega}\sum_{\vec{p}}\sum_{ \sigma',\sigma'\neq\sigma}\gamma'_{1} (\vec{p},\vec{k})n_{\sigma'}(\vec{p},t)n_{\sigma}(\vec{k},t),\\
&= -\bar{\gamma}_1  \int_{0}^{\infty} 
p^2(k^2+p^2) n_{\sigma}(k,t)n_{\sigma'}(p,t)dp,
\end{align}
where $\bar{\gamma_1}=\frac{3\gamma g^2  }{ 4\pi^3(\Delta^2+\gamma^2/4)}.$ In the last equation, we have used the fact that the initial momentum distribution is spherical symmetrical and carried out the angular integration on the spherical harmonics. The integrand is monotonically increasing  in the momentum $k$ which means a stronger depletion at the higher energy particles. Besides, the $p$ and $k$ dependences in the integrand explains why a system with higher initial temperature shows a much faster decay in the particle densities. Since, at high temperature, thermal fluctuation leads to particles occupying higher energies with momentum $p$ that contributes to much higher decay rate to the particles with momentum $k$. 
Figure~\ref{fig:2bdloss_p_m} (a) and (b) are the evolution of the momentum distribution at low and high temperatures. From the low temperature result in (a), we note the depletion of particles is the strongest near the Fermi Surface as expected from the momentum dependence in eq.~\eqref{eq:micro_eq} where particles with higher momentum have larger decay rate. At low temperatures, the low energy particles experience a stronger decay than their high temperature counterparts in panel (b). This is because the higher occupation in the low momentum part of the distribution at low temperature enters in eq.~\eqref{eq:micro_eq} causing a stronger decay.
 

\section{conclusion}\label{sec:fin}
In conclusion, we provide a microscopic description on the two-body loss rate dynamics of cold atoms near $s$ and $p$ wave Feshbach resonances for a two component system. We derived the loss rate equation from a microscopic theory that includes the microscopic mechanism of two body loss. The loss rate equation is generally an integral-differential equation of the momentum distribution. The loss coefficient is a function that contains the information of the microscopic loss mechanism  and would lead to  the difference between the microscopic theory and the phenomenological equation for an quantum-degenerate gas.
For $s$-wave resonance, the microscopic loss rate equation is equivalent to the phenomenological equation. However, this is because of the vanishing momentum dependence within our leading order calculation. For $p$-wave resonance of an quantum degenerate gas, we observe different loss dynamics from the phenomenological results.  For a gas at high temperatures, $T\gtrsim T_F$, the microscopic information in the loss coefficient is smeared out by thermal average which leads to the resemblance to the phenomenological equation.

\section*{Acknowlegements}
This work has been supported by the Agencia Estatal de Investigación del Ministerio de Ciencia e Innovación (Spain) through Grants No. PID2020-120614GB-I00/AEI/10.13039/501100011033 (ENACT). CHH acknowledges the PhD fellowship from DIPC and M.A.Cazalilla for interesting discussions.

\appendix
\section{Derivation of the loss rate equation}\label{app:A}

The momentum distribution $n_{\sigma}(\vec{r},t)$ for particle with momentum $\vec{r}$ at  time $t$ using perturbation expansion to leading order is
\begin{align}\label{eq:m}
&n_{ \sigma }(\vec{r},t) - n^0_\sigma(\vec{r}) \simeq i \langle \bar{c}_{\vec{r}\sigma,-}(t) c_{\vec{r}\sigma,+}(t)  \mathcal{L}\rangle_{0},\notag\\
&\notag\\
&= \frac{-i}{2\Omega}\sum_{\vec{pkq},\alpha\beta,mn} Q_{\alpha\beta}\biggl\{ \int_{-\infty}^t \Theta(t_1)dt_1\,    \sigma^3_{mn}\left[ \delta U_{l}(\vec{p},\vec{k},\vec{q})\right]\notag  \\
&\qquad\qquad\qquad\qquad\times O^{mn}_{\sigma\alpha\beta}(t,t_1;\vec{r},\vec{p},\vec{k},\vec{q})\notag\\
&-i \int_{-\infty}^t \Theta(t_1)dt_1\, \sigma^0_{mn}\gamma'_{l}(\vec{p},\vec{k},\vec{q})  O^{mn}_{\sigma\alpha\beta}(t,t_1;\vec{r},\vec{p},\vec{k},\vec{q})\notag\\
&+2i   \int_{-\infty}^t \Theta(t_1)dt_1 \,  \gamma'_{l}(\vec{p},\vec{k},\vec{q}) O^{-+}_{\sigma\alpha\beta}(t,t_1;\vec{r},\vec{p},\vec{k},\vec{q})
\biggr\},
\end{align}
where  $n^0_\sigma(\vec{r}) = \langle c^\dag_{\vec{r}\sigma,-}(t) c_{\vec{r}\sigma,+}(t) \rangle_{0} $ is the momentum distribution of the  fermions with spin $\sigma$ in the initial state described by the (non-interacting) Fermi-Dirac distribution with initial temperature $T$. In the above expression $\langle \ldots \rangle_0$ stands for
Keldysh time-ordered average with the weight $e^{iS_0}$ in eq~\eqref{eq:fesh2}. $m,n=\pm$ stands for the time ordered and anti-time ordered segment of the Keldysh contour. In addition, we have also introduced the following notation for the six fermion-operator expectation values:
%
\begin{align}
&O^{mn}_{\sigma\alpha\beta}(t,t_1;\vec{r},\vec{p},\vec{k},\vec{q})\notag \\
&=\langle \bar{c}_{\vec{r}\sigma,-}(t) c_{\vec{r}\sigma,+}(t)\, \bar{c}_{\vec{p}\alpha,m}(t_1) \bar{c}_{\vec{k}\beta,m}(t_1)c_{\vec{k+q}\beta,n}(t_1)c_{\vec{p-q}\alpha,n}(t_1)  \rangle_{0}. 
 \end{align}
 %
Expanding  the sum including $\sigma^3_{mn}$ and $\sigma^0_{mn}$, we have
\begin{align}\label{eq:O30}
\sum_{mn} &\sigma^3_{mn}O^{mn}_{\sigma\alpha\beta}(t,t_1;\vec{r},\vec{p},\vec{k},\vec{q}) \notag\\
&=O^{++}_{\sigma\alpha\beta}(t,t_1;\vec{r},\vec{p},\vec{k},\vec{q}) -O^{--}_{\sigma\alpha\beta}(t,t_1;\vec{r},\vec{p},\vec{k},\vec{q}) .\\
\sum_{mn}&\sigma^0_{mn} O^{mn}_{\sigma\alpha\beta}(t,t_1;\vec{r},\vec{p},\vec{k},\vec{q}) \notag\\
&=O^{++}_{\sigma\alpha\beta}(t,t_1;\vec{r},\vec{p},\vec{k},\vec{q}) +O^{--}_{\sigma\alpha\beta}(t,t_1;\vec{r},\vec{p},\vec{k},\vec{q}) .
\end{align}
Next, applying Wick theorem yields:
\begin{align}
& O^{mn}_{\sigma\alpha\beta}(t,t_1;\vec{r},\vec{p},\vec{k},\vec{q})\notag \\
&= \biggl [ \langle \bar{c}_{\vec{k}\beta,m}(t_1)c_{\vec{k+q}\beta,n}(t_1)  \rangle_{0}\langle c_{\vec{r}\sigma,+}(t)\, \bar{c}_{\vec{p}\alpha,m}(t_1) \rangle_{0}
\notag\\& \qquad\qquad\qquad\qquad\qquad\qquad\times\langle \bar{c}_{\vec{r}\sigma,-}(t)c_{\vec{p-q}\alpha,n}(t_1)\rangle_{0}\notag \\ 
&+
\langle  \bar{c}_{\vec{p}\alpha,m}(t_1)c_{\vec{p-q}\alpha,n}(t_1)  \rangle_{0}\langle c_{\vec{r}\sigma,+}(t)\, \bar{c}_{\vec{k}\beta,m}(t_1) \rangle_{0}\notag\\& \qquad\qquad\qquad\qquad\qquad\qquad\times\langle \bar{c}_{\vec{r}\sigma,-}(t)c_{\vec{k+q}\beta,n}(t_1)  \rangle_{0} 
\biggr]\notag \\
&-  \biggl[ \langle \bar{c}_{\vec{k}\beta,m}(t_1)c_{\vec{p-q}\alpha,n}(t_1)  \rangle_{0}\langle c_{\vec{r}\sigma,+}(t)\, \bar{c}_{\vec{p}\alpha,m}(t_1) \rangle_{0}\notag\\& \qquad\qquad\qquad\qquad\qquad\qquad\times\langle \bar{c}_{\vec{r}\sigma,-}(t)c_{\vec{k+q}\beta,n}(t_1)  \rangle_{0}\notag \\
&+\langle \bar{c}_{\vec{p}\alpha,m}(t_1)c_{\vec{k+q}\beta,n}(t_1)  \rangle_{0}\langle c_{\vec{r}\sigma,+}(t)\, \bar{c}_{\vec{k}\beta,m}(t_1) \rangle_{0}\notag\\& \qquad\qquad\qquad\qquad\qquad\qquad\times\langle \bar{c}_{\vec{r}\sigma,-}(t)c_{\vec{p-q}\alpha,n}(t_1)  \rangle_{0}
\biggr],
\end{align}
which yields
\begin{align}\label{eq:OPPMM}
&O^{++}_{\sigma\alpha\beta}(t,t_1;\vec{r},\vec{p},\vec{k},\vec{q})=O^{--}_{\sigma\alpha\beta}(t,t_1;\vec{r},\vec{p},\vec{k},\vec{q})\notag\\
&=\biggl[   \delta_{\vec{q,0}} \delta_{\vec{r,p}}\delta_{\sigma,\alpha} n_{\beta}^0(\vec{k})   + \delta_{\vec{q,0}} \delta_{\vec{r,k}}\delta_{\sigma,\beta} n_\alpha^0(\vec{p})   \notag  \\
&-\delta_{\vec{q,p-k}} \delta_{\vec{r,p}} \delta_{\sigma,\alpha}\delta_{\sigma,\beta}n_{\beta}^0(\vec{k})   - \delta_{\vec{q,p-k}} \delta_{\vec{r,k}} \delta_{\sigma,\alpha}\delta_{\sigma,\beta} n_\alpha^0(\vec{p})   \biggr ] \notag \\
&\times \biggl[\tilde{\theta}(t-t_1) \big(1-n_{\sigma}^0(\vec{r}) \big)n_{\sigma}^0(\vec{r}) - \theta(t_1-t) \left[n_{\sigma}^0(\vec{r})\right]^2\biggr].
\end{align}
 and
 \begin{align}\label{eq:OMP}
&O^{-+}_{\sigma\alpha\beta}(t,t_1;\vec{r},\vec{p},\vec{k},\vec{q}) = -\biggl[   \delta_{\vec{q,0}} \delta_{\vec{r,p}} \delta_{\alpha,\sigma} n_\beta^0(\vec{k})+ \delta_{\vec{q,0}} \delta_{\vec{r,k}} \delta_{\beta,\sigma}n_\alpha^0(\vec{p})   \notag\\
&- \delta_{\vec{q,p-k}} \delta_{\vec{r,p}} \delta_{\alpha,\sigma}\delta_{\beta,\sigma}n_\beta^0(\vec{k})  -\delta_{\vec{q,p-k}} \delta_{\vec{r,k}}\delta_{\alpha,\sigma}\delta_{\beta,\sigma} n_\alpha^0(\vec{p})\biggr ] \left[n_\sigma^0(\vec{r})\right]^2.
  \end{align}
  Note that the exponential phase dependence on  $t$ and $t_1$ is cancelled in the above first order expectation values. Finally, combining Eq.~\eqref{eq:m}, \eqref{eq:O30}, \eqref{eq:OPPMM} with Eq.~\eqref{eq:OMP} and re-arranging the momentum and spin indices yields:
 \begin{align} 
n_\sigma(\vec{r},t)&-n_\sigma^0(\vec{r}) = -\frac{2}{\Omega}\int_{-\infty}^t\Theta(t_1) dt_1 \notag \\
&\times \sum_{\vec{k}\beta}Q_{\sigma\beta}\left[\gamma'_{l}(\vec{r,k,0})-\delta_{\sigma\beta}\gamma'_{l}(\vec{r,k,r-k})\right] n^0_{\sigma}(\vec{r})n^0_{\beta}(\vec{k}).
\end{align}
Taking $t\to 0$,  we arrive at the loss rate equation given in 
Eq.~\eqref{eq:micro}.

\bibliography{note_OFR_loss,lindblad}

\begin{thebibliography}{28}%
\makeatletter
\providecommand \@ifxundefined [1]{%
 \@ifx{#1\undefined}
}%
\providecommand \@ifnum [1]{%
 \ifnum #1\expandafter \@firstoftwo
 \else \expandafter \@secondoftwo
 \fi
}%
\providecommand \@ifx [1]{%
 \ifx #1\expandafter \@firstoftwo
 \else \expandafter \@secondoftwo
 \fi
}%
\providecommand \natexlab [1]{#1}%
\providecommand \enquote  [1]{``#1''}%
\providecommand \bibnamefont  [1]{#1}%
\providecommand \bibfnamefont [1]{#1}%
\providecommand \citenamefont [1]{#1}%
\providecommand \href@noop [0]{\@secondoftwo}%
\providecommand \href [0]{\begingroup \@sanitize@url \@href}%
\providecommand \@href[1]{\@@startlink{#1}\@@href}%
\providecommand \@@href[1]{\endgroup#1\@@endlink}%
\providecommand \@sanitize@url [0]{\catcode `\\12\catcode `\$12\catcode
  `\&12\catcode `\#12\catcode `\^12\catcode `\_12\catcode `\%12\relax}%
\providecommand \@@startlink[1]{}%
\providecommand \@@endlink[0]{}%
\providecommand \url  [0]{\begingroup\@sanitize@url \@url }%
\providecommand \@url [1]{\endgroup\@href {#1}{\urlprefix }}%
\providecommand \urlprefix  [0]{URL }%
\providecommand \Eprint [0]{\href }%
\providecommand \doibase [0]{https://doi.org/}%
\providecommand \selectlanguage [0]{\@gobble}%
\providecommand \bibinfo  [0]{\@secondoftwo}%
\providecommand \bibfield  [0]{\@secondoftwo}%
\providecommand \translation [1]{[#1]}%
\providecommand \BibitemOpen [0]{}%
\providecommand \bibitemStop [0]{}%
\providecommand \bibitemNoStop [0]{.\EOS\space}%
\providecommand \EOS [0]{\spacefactor3000\relax}%
\providecommand \BibitemShut  [1]{\csname bibitem#1\endcsname}%
\let\auto@bib@innerbib\@empty
\bibitem [{\citenamefont {Chin}\ \emph
  {et~al.}(2010{\natexlab{a}})\citenamefont {Chin}, \citenamefont {Grimm},
  \citenamefont {Julienne},\ and\ \citenamefont
  {Tiesinga}}]{RevModPhys.82.1225_CC}%
  \BibitemOpen
  \bibfield  {author} {\bibinfo {author} {\bibfnamefont {C.}~\bibnamefont
  {Chin}}, \bibinfo {author} {\bibfnamefont {R.}~\bibnamefont {Grimm}},
  \bibinfo {author} {\bibfnamefont {P.}~\bibnamefont {Julienne}},\ and\
  \bibinfo {author} {\bibfnamefont {E.}~\bibnamefont {Tiesinga}},\ }\bibfield
  {title} {\bibinfo {title} {Feshbach resonances in ultracold gases},\ }\href
  {https://doi.org/10.1103/RevModPhys.82.1225} {\bibfield  {journal} {\bibinfo
  {journal} {Rev. Mod. Phys.}\ }\textbf {\bibinfo {volume} {82}},\ \bibinfo
  {pages} {1225} (\bibinfo {year} {2010}{\natexlab{a}})}\BibitemShut {NoStop}%
\bibitem [{\citenamefont {Duine}\ and\ \citenamefont
  {Stoof}(2004)}]{DUINE2004115_review}%
  \BibitemOpen
  \bibfield  {author} {\bibinfo {author} {\bibfnamefont {R.}~\bibnamefont
  {Duine}}\ and\ \bibinfo {author} {\bibfnamefont {H.}~\bibnamefont {Stoof}},\
  }\bibfield  {title} {\bibinfo {title} {Atom–molecule coherence in bose
  gases},\ }\href
  {https://doi.org/https://doi.org/10.1016/j.physrep.2004.03.003} {\bibfield
  {journal} {\bibinfo  {journal} {Physics Reports}\ }\textbf {\bibinfo {volume}
  {396}},\ \bibinfo {pages} {115} (\bibinfo {year} {2004})}\BibitemShut
  {NoStop}%
\bibitem [{\citenamefont {Timmermans}\ \emph {et~al.}(1999)\citenamefont
  {Timmermans}, \citenamefont {Tommasini}, \citenamefont {Hussein},\ and\
  \citenamefont {Kerman}}]{TIMMERMANS1999199_review}%
  \BibitemOpen
  \bibfield  {author} {\bibinfo {author} {\bibfnamefont {E.}~\bibnamefont
  {Timmermans}}, \bibinfo {author} {\bibfnamefont {P.}~\bibnamefont
  {Tommasini}}, \bibinfo {author} {\bibfnamefont {M.}~\bibnamefont {Hussein}},\
  and\ \bibinfo {author} {\bibfnamefont {A.}~\bibnamefont {Kerman}},\
  }\bibfield  {title} {\bibinfo {title} {Feshbach resonances in atomic
  bose--einstein condensates},\ }\href
  {https://doi.org/https://doi.org/10.1016/S0370-1573(99)00025-3} {\bibfield
  {journal} {\bibinfo  {journal} {Physics Reports}\ }\textbf {\bibinfo {volume}
  {315}},\ \bibinfo {pages} {199} (\bibinfo {year} {1999})}\BibitemShut
  {NoStop}%
\bibitem [{\citenamefont {Jones}\ \emph {et~al.}(2006)\citenamefont {Jones},
  \citenamefont {Tiesinga}, \citenamefont {Lett},\ and\ \citenamefont
  {Julienne}}]{RevModPhys.78.483_review}%
  \BibitemOpen
  \bibfield  {author} {\bibinfo {author} {\bibfnamefont {K.~M.}\ \bibnamefont
  {Jones}}, \bibinfo {author} {\bibfnamefont {E.}~\bibnamefont {Tiesinga}},
  \bibinfo {author} {\bibfnamefont {P.~D.}\ \bibnamefont {Lett}},\ and\
  \bibinfo {author} {\bibfnamefont {P.~S.}\ \bibnamefont {Julienne}},\
  }\bibfield  {title} {\bibinfo {title} {Ultracold photoassociation
  spectroscopy: Long-range molecules and atomic scattering},\ }\href
  {https://doi.org/10.1103/RevModPhys.78.483} {\bibfield  {journal} {\bibinfo
  {journal} {Rev. Mod. Phys.}\ }\textbf {\bibinfo {volume} {78}},\ \bibinfo
  {pages} {483} (\bibinfo {year} {2006})}\BibitemShut {NoStop}%
\bibitem [{\citenamefont {K\"ohler}\ \emph {et~al.}(2006)\citenamefont
  {K\"ohler}, \citenamefont {G\'oral},\ and\ \citenamefont
  {Julienne}}]{RevModPhys.78.1311_review}%
  \BibitemOpen
  \bibfield  {author} {\bibinfo {author} {\bibfnamefont {T.}~\bibnamefont
  {K\"ohler}}, \bibinfo {author} {\bibfnamefont {K.}~\bibnamefont {G\'oral}},\
  and\ \bibinfo {author} {\bibfnamefont {P.~S.}\ \bibnamefont {Julienne}},\
  }\bibfield  {title} {\bibinfo {title} {Production of cold molecules via
  magnetically tunable feshbach resonances},\ }\href
  {https://doi.org/10.1103/RevModPhys.78.1311} {\bibfield  {journal} {\bibinfo
  {journal} {Rev. Mod. Phys.}\ }\textbf {\bibinfo {volume} {78}},\ \bibinfo
  {pages} {1311} (\bibinfo {year} {2006})}\BibitemShut {NoStop}%
\bibitem [{\citenamefont {Bohn}\ and\ \citenamefont
  {Julienne}(1997)}]{Bohn_PhysRevA.56.1486}%
  \BibitemOpen
  \bibfield  {author} {\bibinfo {author} {\bibfnamefont {J.~L.}\ \bibnamefont
  {Bohn}}\ and\ \bibinfo {author} {\bibfnamefont {P.~S.}\ \bibnamefont
  {Julienne}},\ }\bibfield  {title} {\bibinfo {title} {Prospects for
  influencing scattering lengths with far-off-resonant light},\ }\href
  {https://doi.org/10.1103/PhysRevA.56.1486} {\bibfield  {journal} {\bibinfo
  {journal} {Phys. Rev. A}\ }\textbf {\bibinfo {volume} {56}},\ \bibinfo
  {pages} {1486} (\bibinfo {year} {1997})}\BibitemShut {NoStop}%
\bibitem [{\citenamefont {Braaten}\ \emph {et~al.}(2017)\citenamefont
  {Braaten}, \citenamefont {Hammer},\ and\ \citenamefont
  {Lepage}}]{PhysRevA.95.012708_Braaten}%
  \BibitemOpen
  \bibfield  {author} {\bibinfo {author} {\bibfnamefont {E.}~\bibnamefont
  {Braaten}}, \bibinfo {author} {\bibfnamefont {H.-W.}\ \bibnamefont
  {Hammer}},\ and\ \bibinfo {author} {\bibfnamefont {G.~P.}\ \bibnamefont
  {Lepage}},\ }\bibfield  {title} {\bibinfo {title} {Lindblad equation for the
  inelastic loss of ultracold atoms},\ }\href
  {https://doi.org/10.1103/PhysRevA.95.012708} {\bibfield  {journal} {\bibinfo
  {journal} {Phys. Rev. A}\ }\textbf {\bibinfo {volume} {95}},\ \bibinfo
  {pages} {012708} (\bibinfo {year} {2017})}\BibitemShut {NoStop}%
\bibitem [{\citenamefont {Yamazaki}\ \emph {et~al.}(2010)\citenamefont
  {Yamazaki}, \citenamefont {Taie}, \citenamefont {Sugawa},\ and\ \citenamefont
  {Takahashi}}]{PhysRevLett.105.050405_OFR0}%
  \BibitemOpen
  \bibfield  {author} {\bibinfo {author} {\bibfnamefont {R.}~\bibnamefont
  {Yamazaki}}, \bibinfo {author} {\bibfnamefont {S.}~\bibnamefont {Taie}},
  \bibinfo {author} {\bibfnamefont {S.}~\bibnamefont {Sugawa}},\ and\ \bibinfo
  {author} {\bibfnamefont {Y.}~\bibnamefont {Takahashi}},\ }\bibfield  {title}
  {\bibinfo {title} {Submicron spatial modulation of an interatomic interaction
  in a bose-einstein condensate},\ }\href
  {https://doi.org/10.1103/PhysRevLett.105.050405} {\bibfield  {journal}
  {\bibinfo  {journal} {Phys. Rev. Lett.}\ }\textbf {\bibinfo {volume} {105}},\
  \bibinfo {pages} {050405} (\bibinfo {year} {2010})}\BibitemShut {NoStop}%
\bibitem [{\citenamefont {Blatt}\ \emph {et~al.}(2011)\citenamefont {Blatt},
  \citenamefont {Nicholson}, \citenamefont {Bloom}, \citenamefont {Williams},
  \citenamefont {Thomsen}, \citenamefont {Julienne},\ and\ \citenamefont
  {Ye}}]{Blatt_PhysRevLett_2011}%
  \BibitemOpen
  \bibfield  {author} {\bibinfo {author} {\bibfnamefont {S.}~\bibnamefont
  {Blatt}}, \bibinfo {author} {\bibfnamefont {T.~L.}\ \bibnamefont
  {Nicholson}}, \bibinfo {author} {\bibfnamefont {B.~J.}\ \bibnamefont
  {Bloom}}, \bibinfo {author} {\bibfnamefont {J.~R.}\ \bibnamefont {Williams}},
  \bibinfo {author} {\bibfnamefont {J.~W.}\ \bibnamefont {Thomsen}}, \bibinfo
  {author} {\bibfnamefont {P.~S.}\ \bibnamefont {Julienne}},\ and\ \bibinfo
  {author} {\bibfnamefont {J.}~\bibnamefont {Ye}},\ }\bibfield  {title}
  {\bibinfo {title} {Measurement of optical feshbach resonances in an ideal
  gas},\ }\href {https://doi.org/10.1103/PhysRevLett.107.073202} {\bibfield
  {journal} {\bibinfo  {journal} {Phys. Rev. Lett.}\ }\textbf {\bibinfo
  {volume} {107}},\ \bibinfo {pages} {073202} (\bibinfo {year}
  {2011})}\BibitemShut {NoStop}%
\bibitem [{\citenamefont {Ciury\l{}o}\ \emph {et~al.}(2004)\citenamefont
  {Ciury\l{}o}, \citenamefont {Tiesinga}, \citenamefont {Kotochigova},\ and\
  \citenamefont {Julienne}}]{Ciurylo_PhysRevA_2004}%
  \BibitemOpen
  \bibfield  {author} {\bibinfo {author} {\bibfnamefont {R.}~\bibnamefont
  {Ciury\l{}o}}, \bibinfo {author} {\bibfnamefont {E.}~\bibnamefont
  {Tiesinga}}, \bibinfo {author} {\bibfnamefont {S.}~\bibnamefont
  {Kotochigova}},\ and\ \bibinfo {author} {\bibfnamefont {P.~S.}\ \bibnamefont
  {Julienne}},\ }\bibfield  {title} {\bibinfo {title} {Photoassociation
  spectroscopy of cold alkaline-earth-metal atoms near the intercombination
  line},\ }\href {https://doi.org/10.1103/PhysRevA.70.062710} {\bibfield
  {journal} {\bibinfo  {journal} {Phys. Rev. A}\ }\textbf {\bibinfo {volume}
  {70}},\ \bibinfo {pages} {062710} (\bibinfo {year} {2004})}\BibitemShut
  {NoStop}%
\bibitem [{\citenamefont {Enomoto}\ \emph {et~al.}(2008)\citenamefont
  {Enomoto}, \citenamefont {Kasa}, \citenamefont {Kitagawa},\ and\
  \citenamefont {Takahashi}}]{enomoto_PhysRevLett.101.203201}%
  \BibitemOpen
  \bibfield  {author} {\bibinfo {author} {\bibfnamefont {K.}~\bibnamefont
  {Enomoto}}, \bibinfo {author} {\bibfnamefont {K.}~\bibnamefont {Kasa}},
  \bibinfo {author} {\bibfnamefont {M.}~\bibnamefont {Kitagawa}},\ and\
  \bibinfo {author} {\bibfnamefont {Y.}~\bibnamefont {Takahashi}},\ }\bibfield
  {title} {\bibinfo {title} {Optical feshbach resonance using the
  intercombination transition},\ }\href
  {https://doi.org/10.1103/PhysRevLett.101.203201} {\bibfield  {journal}
  {\bibinfo  {journal} {Phys. Rev. Lett.}\ }\textbf {\bibinfo {volume} {101}},\
  \bibinfo {pages} {203201} (\bibinfo {year} {2008})}\BibitemShut {NoStop}%
\bibitem [{\citenamefont {Theis}\ \emph
  {et~al.}(2004{\natexlab{a}})\citenamefont {Theis}, \citenamefont
  {Thalhammer}, \citenamefont {Winkler}, \citenamefont {Hellwig}, \citenamefont
  {Ruff}, \citenamefont {Grimm},\ and\ \citenamefont
  {Denschlag}}]{TheisDenschlag2004}%
  \BibitemOpen
  \bibfield  {author} {\bibinfo {author} {\bibfnamefont {M.}~\bibnamefont
  {Theis}}, \bibinfo {author} {\bibfnamefont {G.}~\bibnamefont {Thalhammer}},
  \bibinfo {author} {\bibfnamefont {K.}~\bibnamefont {Winkler}}, \bibinfo
  {author} {\bibfnamefont {M.}~\bibnamefont {Hellwig}}, \bibinfo {author}
  {\bibfnamefont {G.}~\bibnamefont {Ruff}}, \bibinfo {author} {\bibfnamefont
  {R.}~\bibnamefont {Grimm}},\ and\ \bibinfo {author} {\bibfnamefont {J.~H.}\
  \bibnamefont {Denschlag}},\ }\bibfield  {title} {\bibinfo {title} {Tuning the
  scattering length with an optically induced {F}eshbach resonance},\
  }\href@noop {} {\bibfield  {journal} {\bibinfo  {journal} {Phys.~ Rev.~
  Lett.}\ ,\ \bibinfo {pages} {123001}} (\bibinfo {year}
  {2004}{\natexlab{a}})}\BibitemShut {NoStop}%
\bibitem [{\citenamefont {Goyal}\ \emph {et~al.}(2010)\citenamefont {Goyal},
  \citenamefont {Reichenbach},\ and\ \citenamefont
  {Deutsch}}]{PhysRevA.82.062704_pwave}%
  \BibitemOpen
  \bibfield  {author} {\bibinfo {author} {\bibfnamefont {K.}~\bibnamefont
  {Goyal}}, \bibinfo {author} {\bibfnamefont {I.}~\bibnamefont {Reichenbach}},\
  and\ \bibinfo {author} {\bibfnamefont {I.}~\bibnamefont {Deutsch}},\
  }\bibfield  {title} {\bibinfo {title} {$p$-wave optical feshbach resonances
  in $^{171}\mathrm{Yb}$},\ }\href {https://doi.org/10.1103/PhysRevA.82.062704}
  {\bibfield  {journal} {\bibinfo  {journal} {Phys. Rev. A}\ }\textbf {\bibinfo
  {volume} {82}},\ \bibinfo {pages} {062704} (\bibinfo {year}
  {2010})}\BibitemShut {NoStop}%
\bibitem [{\citenamefont {Yamazaki}\ \emph {et~al.}(2013)\citenamefont
  {Yamazaki}, \citenamefont {Taie}, \citenamefont {Sugawa}, \citenamefont
  {Enomoto},\ and\ \citenamefont {Takahashi}}]{Yamazaki_PhysRevA.87.010704}%
  \BibitemOpen
  \bibfield  {author} {\bibinfo {author} {\bibfnamefont {R.}~\bibnamefont
  {Yamazaki}}, \bibinfo {author} {\bibfnamefont {S.}~\bibnamefont {Taie}},
  \bibinfo {author} {\bibfnamefont {S.}~\bibnamefont {Sugawa}}, \bibinfo
  {author} {\bibfnamefont {K.}~\bibnamefont {Enomoto}},\ and\ \bibinfo {author}
  {\bibfnamefont {Y.}~\bibnamefont {Takahashi}},\ }\bibfield  {title} {\bibinfo
  {title} {Observation of a $p$-wave optical feshbach resonance},\ }\href
  {https://doi.org/10.1103/PhysRevA.87.010704} {\bibfield  {journal} {\bibinfo
  {journal} {Phys. Rev. A}\ }\textbf {\bibinfo {volume} {87}},\ \bibinfo
  {pages} {010704} (\bibinfo {year} {2013})}\BibitemShut {NoStop}%
\bibitem [{\citenamefont {Huang}\ \emph {et~al.}(2023)\citenamefont {Huang},
  \citenamefont {Giamarchi},\ and\ \citenamefont
  {Cazalilla}}]{PhysRevResearch.5.043192_CH}%
  \BibitemOpen
  \bibfield  {author} {\bibinfo {author} {\bibfnamefont {C.-H.}\ \bibnamefont
  {Huang}}, \bibinfo {author} {\bibfnamefont {T.}~\bibnamefont {Giamarchi}},\
  and\ \bibinfo {author} {\bibfnamefont {M.~A.}\ \bibnamefont {Cazalilla}},\
  }\bibfield  {title} {\bibinfo {title} {Modeling particle loss in open systems
  using keldysh path integral and second order cumulant expansion},\ }\href
  {https://doi.org/10.1103/PhysRevResearch.5.043192} {\bibfield  {journal}
  {\bibinfo  {journal} {Phys. Rev. Res.}\ }\textbf {\bibinfo {volume} {5}},\
  \bibinfo {pages} {043192} (\bibinfo {year} {2023})}\BibitemShut {NoStop}%
\bibitem [{\citenamefont {Rossini}\ \emph {et~al.}(2021)\citenamefont
  {Rossini}, \citenamefont {Ghermaoui}, \citenamefont {Aguilera}, \citenamefont
  {Vatr\'e}, \citenamefont {Bouganne}, \citenamefont {Beugnon}, \citenamefont
  {Gerbier},\ and\ \citenamefont {Mazza}}]{Rossini_PhysRevA.103.L060201}%
  \BibitemOpen
  \bibfield  {author} {\bibinfo {author} {\bibfnamefont {D.}~\bibnamefont
  {Rossini}}, \bibinfo {author} {\bibfnamefont {A.}~\bibnamefont {Ghermaoui}},
  \bibinfo {author} {\bibfnamefont {M.~B.}\ \bibnamefont {Aguilera}}, \bibinfo
  {author} {\bibfnamefont {R.}~\bibnamefont {Vatr\'e}}, \bibinfo {author}
  {\bibfnamefont {R.}~\bibnamefont {Bouganne}}, \bibinfo {author}
  {\bibfnamefont {J.}~\bibnamefont {Beugnon}}, \bibinfo {author} {\bibfnamefont
  {F.}~\bibnamefont {Gerbier}},\ and\ \bibinfo {author} {\bibfnamefont
  {L.}~\bibnamefont {Mazza}},\ }\bibfield  {title} {\bibinfo {title} {Strong
  correlations in lossy one-dimensional quantum gases: From the quantum zeno
  effect to the generalized gibbs ensemble},\ }\href
  {https://doi.org/10.1103/PhysRevA.103.L060201} {\bibfield  {journal}
  {\bibinfo  {journal} {Phys. Rev. A}\ }\textbf {\bibinfo {volume} {103}},\
  \bibinfo {pages} {L060201} (\bibinfo {year} {2021})}\BibitemShut {NoStop}%
\bibitem [{\citenamefont {Tomita}\ \emph {et~al.}(2019)\citenamefont {Tomita},
  \citenamefont {Nakajima}, \citenamefont {Takasu},\ and\ \citenamefont
  {Takahashi}}]{Tomita_PhysRevA_2019}%
  \BibitemOpen
  \bibfield  {author} {\bibinfo {author} {\bibfnamefont {T.}~\bibnamefont
  {Tomita}}, \bibinfo {author} {\bibfnamefont {S.}~\bibnamefont {Nakajima}},
  \bibinfo {author} {\bibfnamefont {Y.}~\bibnamefont {Takasu}},\ and\ \bibinfo
  {author} {\bibfnamefont {Y.}~\bibnamefont {Takahashi}},\ }\bibfield  {title}
  {\bibinfo {title} {Dissipative bose-hubbard system with intrinsic two-body
  loss},\ }\href {https://doi.org/10.1103/PhysRevA.99.031601} {\bibfield
  {journal} {\bibinfo  {journal} {Phys. Rev. A}\ }\textbf {\bibinfo {volume}
  {99}},\ \bibinfo {pages} {031601} (\bibinfo {year} {2019})}\BibitemShut
  {NoStop}%
\bibitem [{\citenamefont {Syassen}\ \emph {et~al.}(2008)\citenamefont
  {Syassen}, \citenamefont {Bauer}, \citenamefont {Lettner}, \citenamefont
  {Volz}, \citenamefont {Dietze}, \citenamefont {Garc{\'\i}a-Ripoll},
  \citenamefont {Cirac}, \citenamefont {Rempe},\ and\ \citenamefont
  {D{\"u}rr}}]{Syassen_science_2008}%
  \BibitemOpen
  \bibfield  {author} {\bibinfo {author} {\bibfnamefont {N.}~\bibnamefont
  {Syassen}}, \bibinfo {author} {\bibfnamefont {D.~M.}\ \bibnamefont {Bauer}},
  \bibinfo {author} {\bibfnamefont {M.}~\bibnamefont {Lettner}}, \bibinfo
  {author} {\bibfnamefont {T.}~\bibnamefont {Volz}}, \bibinfo {author}
  {\bibfnamefont {D.}~\bibnamefont {Dietze}}, \bibinfo {author} {\bibfnamefont
  {J.~J.}\ \bibnamefont {Garc{\'\i}a-Ripoll}}, \bibinfo {author} {\bibfnamefont
  {J.~I.}\ \bibnamefont {Cirac}}, \bibinfo {author} {\bibfnamefont
  {G.}~\bibnamefont {Rempe}},\ and\ \bibinfo {author} {\bibfnamefont
  {S.}~\bibnamefont {D{\"u}rr}},\ }\bibfield  {title} {\bibinfo {title} {Strong
  dissipation inhibits losses and induces correlations in cold molecular
  gases},\ }\href {https://doi.org/10.1126/science.1155309} {\bibfield
  {journal} {\bibinfo  {journal} {Science}\ }\textbf {\bibinfo {volume}
  {320}},\ \bibinfo {pages} {1329} (\bibinfo {year} {2008})},\ \Eprint
  {https://arxiv.org/abs/https://www.science.org/doi/pdf/10.1126/science.1155309}
  {https://www.science.org/doi/pdf/10.1126/science.1155309} \BibitemShut
  {NoStop}%
\bibitem [{\citenamefont {Kamenev}(2011)}]{kamenev_2011}%
  \BibitemOpen
  \bibfield  {author} {\bibinfo {author} {\bibfnamefont {A.}~\bibnamefont
  {Kamenev}},\ }\href {https://doi.org/10.1017/CBO9781139003667} {\emph
  {\bibinfo {title} {Field Theory of Non-Equilibrium Systems}}}\ (\bibinfo
  {publisher} {Cambridge University Press},\ \bibinfo {year}
  {2011})\BibitemShut {NoStop}%
\bibitem [{\citenamefont {Sieberer}\ \emph {et~al.}(2016)\citenamefont
  {Sieberer}, \citenamefont {Buchhold},\ and\ \citenamefont
  {Diehl}}]{Sieberer_IOP_2016}%
  \BibitemOpen
  \bibfield  {author} {\bibinfo {author} {\bibfnamefont {L.~M.}\ \bibnamefont
  {Sieberer}}, \bibinfo {author} {\bibfnamefont {M.}~\bibnamefont {Buchhold}},\
  and\ \bibinfo {author} {\bibfnamefont {S.}~\bibnamefont {Diehl}},\ }\bibfield
   {title} {\bibinfo {title} {Keldysh field theory for driven open quantum
  systems},\ }\href {https://doi.org/10.1088/0034-4885/79/9/096001} {\bibfield
  {journal} {\bibinfo  {journal} {Reports on Progress in Physics}\ }\textbf
  {\bibinfo {volume} {79}},\ \bibinfo {pages} {096001} (\bibinfo {year}
  {2016})}\BibitemShut {NoStop}%
\bibitem [{\citenamefont {Kubo}(1980)}]{Kubo1980}%
  \BibitemOpen
  \bibfield  {author} {\bibinfo {author} {\bibfnamefont {K.}~\bibnamefont
  {Kubo}},\ }\bibfield  {title} {\bibinfo {title} {Magnetic susceptibility of
  the strongly correlated {H}ubbard model},\ }\href@noop {} {\bibfield
  {journal} {\bibinfo  {journal} {Prog. Theor. Phys.}\ }\textbf {\bibinfo
  {volume} {64}},\ \bibinfo {pages} {758} (\bibinfo {year} {1980})}\BibitemShut
  {NoStop}%
\bibitem [{\citenamefont {Yoshida}\ and\ \citenamefont
  {Ueda}(2015)}]{PhysRevLett.115.135303_pwave_2}%
  \BibitemOpen
  \bibfield  {author} {\bibinfo {author} {\bibfnamefont {S.~M.}\ \bibnamefont
  {Yoshida}}\ and\ \bibinfo {author} {\bibfnamefont {M.}~\bibnamefont {Ueda}},\
  }\bibfield  {title} {\bibinfo {title} {Universal high-momentum asymptote and
  thermodynamic relations in a spinless fermi gas with a resonant $p$-wave
  interaction},\ }\href {https://doi.org/10.1103/PhysRevLett.115.135303}
  {\bibfield  {journal} {\bibinfo  {journal} {Phys. Rev. Lett.}\ }\textbf
  {\bibinfo {volume} {115}},\ \bibinfo {pages} {135303} (\bibinfo {year}
  {2015})}\BibitemShut {NoStop}%
\bibitem [{\citenamefont {Gubbels}\ and\ \citenamefont
  {Stoof}(2007)}]{PhysRevLett.99.190406_pwave_1}%
  \BibitemOpen
  \bibfield  {author} {\bibinfo {author} {\bibfnamefont {K.~B.}\ \bibnamefont
  {Gubbels}}\ and\ \bibinfo {author} {\bibfnamefont {H.~T.~C.}\ \bibnamefont
  {Stoof}},\ }\bibfield  {title} {\bibinfo {title} {Theory for $p$-wave
  feshbach molecules},\ }\href {https://doi.org/10.1103/PhysRevLett.99.190406}
  {\bibfield  {journal} {\bibinfo  {journal} {Phys. Rev. Lett.}\ }\textbf
  {\bibinfo {volume} {99}},\ \bibinfo {pages} {190406} (\bibinfo {year}
  {2007})}\BibitemShut {NoStop}%
\bibitem [{\citenamefont {Theis}\ \emph
  {et~al.}(2004{\natexlab{b}})\citenamefont {Theis}, \citenamefont
  {Thalhammer}, \citenamefont {Winkler}, \citenamefont {Hellwig}, \citenamefont
  {Ruff}, \citenamefont {Grimm},\ and\ \citenamefont
  {Denschlag}}]{Thies_PhysRevLett.93.123001}%
  \BibitemOpen
  \bibfield  {author} {\bibinfo {author} {\bibfnamefont {M.}~\bibnamefont
  {Theis}}, \bibinfo {author} {\bibfnamefont {G.}~\bibnamefont {Thalhammer}},
  \bibinfo {author} {\bibfnamefont {K.}~\bibnamefont {Winkler}}, \bibinfo
  {author} {\bibfnamefont {M.}~\bibnamefont {Hellwig}}, \bibinfo {author}
  {\bibfnamefont {G.}~\bibnamefont {Ruff}}, \bibinfo {author} {\bibfnamefont
  {R.}~\bibnamefont {Grimm}},\ and\ \bibinfo {author} {\bibfnamefont {J.~H.}\
  \bibnamefont {Denschlag}},\ }\bibfield  {title} {\bibinfo {title} {Tuning the
  scattering length with an optically induced feshbach resonance},\ }\href
  {https://doi.org/10.1103/PhysRevLett.93.123001} {\bibfield  {journal}
  {\bibinfo  {journal} {Phys. Rev. Lett.}\ }\textbf {\bibinfo {volume} {93}},\
  \bibinfo {pages} {123001} (\bibinfo {year} {2004}{\natexlab{b}})}\BibitemShut
  {NoStop}%
\bibitem [{\citenamefont {Napolitano}\ \emph {et~al.}(1994)\citenamefont
  {Napolitano}, \citenamefont {Weiner}, \citenamefont {Williams},\ and\
  \citenamefont {Julienne}}]{Napolitano_PhysRevLett.73.1352}%
  \BibitemOpen
  \bibfield  {author} {\bibinfo {author} {\bibfnamefont {R.}~\bibnamefont
  {Napolitano}}, \bibinfo {author} {\bibfnamefont {J.}~\bibnamefont {Weiner}},
  \bibinfo {author} {\bibfnamefont {C.~J.}\ \bibnamefont {Williams}},\ and\
  \bibinfo {author} {\bibfnamefont {P.~S.}\ \bibnamefont {Julienne}},\
  }\bibfield  {title} {\bibinfo {title} {Line shapes of high resolution
  photoassociation spectra of optically cooled atoms},\ }\href
  {https://doi.org/10.1103/PhysRevLett.73.1352} {\bibfield  {journal} {\bibinfo
   {journal} {Phys. Rev. Lett.}\ }\textbf {\bibinfo {volume} {73}},\ \bibinfo
  {pages} {1352} (\bibinfo {year} {1994})}\BibitemShut {NoStop}%
\bibitem [{\citenamefont {Kim}\ \emph {et~al.}(2016)\citenamefont {Kim},
  \citenamefont {Lee}, \citenamefont {Lee}, \citenamefont {Shin},\ and\
  \citenamefont {Mun}}]{Kim_PhysRevA.94.042703}%
  \BibitemOpen
  \bibfield  {author} {\bibinfo {author} {\bibfnamefont {M.-S.}\ \bibnamefont
  {Kim}}, \bibinfo {author} {\bibfnamefont {J.}~\bibnamefont {Lee}}, \bibinfo
  {author} {\bibfnamefont {J.~H.}\ \bibnamefont {Lee}}, \bibinfo {author}
  {\bibfnamefont {Y.}~\bibnamefont {Shin}},\ and\ \bibinfo {author}
  {\bibfnamefont {J.}~\bibnamefont {Mun}},\ }\bibfield  {title} {\bibinfo
  {title} {Measurements of optical feshbach resonances of $^{174}\mathrm{Yb}$
  atoms},\ }\href {https://doi.org/10.1103/PhysRevA.94.042703} {\bibfield
  {journal} {\bibinfo  {journal} {Phys. Rev. A}\ }\textbf {\bibinfo {volume}
  {94}},\ \bibinfo {pages} {042703} (\bibinfo {year} {2016})}\BibitemShut
  {NoStop}%
\bibitem [{\citenamefont {Chin}\ \emph
  {et~al.}(2010{\natexlab{b}})\citenamefont {Chin}, \citenamefont {Grimm},
  \citenamefont {Julienne},\ and\ \citenamefont
  {Tiesinga}}]{RevModPhys.82.1225}%
  \BibitemOpen
  \bibfield  {author} {\bibinfo {author} {\bibfnamefont {C.}~\bibnamefont
  {Chin}}, \bibinfo {author} {\bibfnamefont {R.}~\bibnamefont {Grimm}},
  \bibinfo {author} {\bibfnamefont {P.}~\bibnamefont {Julienne}},\ and\
  \bibinfo {author} {\bibfnamefont {E.}~\bibnamefont {Tiesinga}},\ }\bibfield
  {title} {\bibinfo {title} {Feshbach resonances in ultracold gases},\ }\href
  {https://doi.org/10.1103/RevModPhys.82.1225} {\bibfield  {journal} {\bibinfo
  {journal} {Rev. Mod. Phys.}\ }\textbf {\bibinfo {volume} {82}},\ \bibinfo
  {pages} {1225} (\bibinfo {year} {2010}{\natexlab{b}})}\BibitemShut {NoStop}%
\bibitem [{\citenamefont {Falco}\ and\ \citenamefont
  {Stoof}(2005)}]{PhysRevA.71.063614_stoof}%
  \BibitemOpen
  \bibfield  {author} {\bibinfo {author} {\bibfnamefont {G.~M.}\ \bibnamefont
  {Falco}}\ and\ \bibinfo {author} {\bibfnamefont {H.~T.~C.}\ \bibnamefont
  {Stoof}},\ }\bibfield  {title} {\bibinfo {title} {Atom-molecule theory of
  broad feshbach resonances},\ }\href
  {https://doi.org/10.1103/PhysRevA.71.063614} {\bibfield  {journal} {\bibinfo
  {journal} {Phys. Rev. A}\ }\textbf {\bibinfo {volume} {71}},\ \bibinfo
  {pages} {063614} (\bibinfo {year} {2005})}\BibitemShut {NoStop}%
\end{thebibliography}%

\end{document}